# Heteroepitaxial growth of Anatase(001) films on SrTiO$_3$(001) by PLD and MBE

A. Crespo[1], J. Gallenberger[2,3], M. De Santis[2], V. Langlais[4], F. Carla[5], J .M. Caicedo[6], J. Rius[1] and X. Torrelles[1]

[1] Institut de Ciència de Materials de Barcelona (ICMAB-CSIC), Campus de la UAB, 08193, Catalonia, Spain
[2] Université Grenoble Alpes, CNRS, Grenoble INP, Institut Néel, 38042 Grenoble, France,
[3] Technische Universität Darmstadt, 64287 Darmstadt, Germany
[4] Centre d'Elaboration de Matériaux et Etudes Structurales (CEMES-CNRS), 31055 Toulouse, France
[5] Diamond Light Source, Harwell Science and Innovation Campus, Didcot, Oxfordshire, OX11 0DE (UK)
[6] Catalan Institute of Nanoscience and Nanotechnology (ICN2), Campus UAB, Building ICN2, Bellaterra, Catalonia, E-0193, Spain

**Abstract**

The epitaxial growth of anatase (001) films deposited by pulsed laser deposition (PLD) and molecular beam epitaxy (MBE) on SrTiO$_3$ (001) (STO) single crystals has been studied using X-ray diffraction and surface sensitivity UHV techniques. The evolution of the strain represented by the microstrain and the change of the in-plane and out-of-plane lattice parameters with film growth temperature, the effect of the annealing temperature and the influence of the oxygen content of the film have been investigated.

The out-of-plane lattice strain shows a compressive (-0.2%) or expansive (+0.3%) behavior, in the range 600 - 900ºC, for temperatures below or above 700ºC, respectively. The in-plane lattice parameters, as well as the cell volume of the film, remain under compression over the entire temperature range explored.

PLD films grow into square islands that align with the surface lattice directions of the STO substrate. The maximum size of these islands is reached at growth temperatures close to 875-925ºC. Film annealing at temperatures of 800°C or higher melts the islands into flat terraces. Larger terraces are reached at high annealing temperatures of 925°C for extended periods of 12 hours. This procedure allows flat surface terrace sizes of up to 650 nm to be achieved.

The crystalline quality achieved in anatase films prepared by PLD or MBE growth methods is similar. The two-step anatase growth process used during the synthesis of the films with both methods: film growth and post-annealing treatment in oxygen or air at ambient pressure, using temperature and time as key parameters, allows to control the surface terrace size and stoichiometry of the films, as well as the anatase/rutile intermixing rates at sufficiently high temperatures. This growth process could allow the substitution of their equivalent single crystals. The range of applicability of these films would include their use as structural and electronic model systems, or in harsh experimental conditions due to their low production cost.

# 1.- INTRODUCTION

Titanium dioxide ($TiO_2$) is a wide band-gap semiconductor, nowadays still considered a very promising material because of its physical and chemical properties and its great variety of applications in different scientific and technological areas of relevance such as photovoltaic conversion of light to electricity [1], photocatalysis [2] or solar cells as electrodes [3]. Among the different polymorphs of $TiO_2$, rutile is the most studied one in terms of both experimental and theoretical fundamental properties, although it is anatase that shows both the best catalytic properties [4] and a number of interesting behaviors [1,5-8].

One aspect that limits the photocatalytic applications of titania is its band gap. Since the absorption edges of rutile and anatase are at ≈ 3.0 eV and ≈ 3.2 eV, respectively, light absorption will only be significant in the UV portion of the spectrum. However, most energy in the solar spectrum is carried by lower energy photons. Consequently, attempts at modifying the absorption edge of titania to increase its efficiency against visible light have been performed either by a partial substitution of oxygen with another electronegative atom or by the absorption of molecular species that can take in visible light and donate electrons or holes to the titania [9-13].

The anatase phase has attracted particular attention due to its longer carrier lifetime, [14] longer diffusion length, [15] and higher electron mobility [16] with respect to the rutile phase. Different facets of anatase-$TiO_2$ exhibit different activities for different reactants. To understand the origin of such morphology-dependent activity, numerous experimental studies have been carried out. Since the pioneering work of [17] on the preparation of a well-defined anatase $TiO_2$ single crystal containing 47% (001) facet, researchers devoted intensive efforts to enhance its efficiency by varying the (001)-to-(101) facet ratio [18]. Theoretical studies studying the mechanisms of $H_2$ dissociation in the three anatase faces (001) (100) and (101) indicate that this process is more favorable in anatase (001) [19]. Most of the works provided insights into the higher photoactivity of the dominant (001) facet for $H_2$ production. In contrast, other studies invoked the low-energy (101) facet for higher photocatalytic activity [20,21]. In the absence of comprehensive fundamental studies pointing out the role of each crystal facet in the photocatalytic redox reactions, the debate still remains unclear and needs to be well clarified.

A second limiting aspect is the scarce knowledge and restricted availability of the anatase-$TiO_2$ (001) face. Oriented rutile substrates, among which the (110) face is the most stable one, can be synthesized in the laboratory and have been exhaustively used as model surfaces in the investigation of fundamental surface phenomena [22]. In contrast, information about (001)-oriented anatase is sparse because: 1) it is difficult to synthesize in the laboratory [23]; 2) it is the most scarce in nature and shows the highest surface energy [24]. Therefore, access to anatase (001) surfaces with large crystal terrace sizes at low cost is of *paramount importance* to advance in the fundamental knowledge of the structural and electronic properties of this surface and for its applications. In the last two decades, the improvements on sample preparation and characterization techniques permitted some experimental studies using well defined anatase surfaces [25-28]. However, the synthesis techniques still need some improvements in order to grow suitable anatase crystals in the laboratory.

The possibility of growing $TiO_2$ over substrates oriented within a specific plane such as $SrTiO_3$ (001) allows the formation of highly ordered films of anatase (001). In. addition, the use of extended synthesis techniques such as PLD (Pulsed Laser Deposition) and MBE (Molecular Beam Epitaxy) could help improve access to this type of material for use in studying its properties, i.e., structural, electronic and photocatalytic.

SrTiO$_3$ (001) (STO) offers one of the best initial conditions for the heteroepitaxial growth of anatase (001). The mismatch between its in-plane parameters is of approx. 3%. On the contrary, substrates like DyScO$_3$ (110), with low surface roughness and with a 4% mismatch between them, produce a mixture of granulated anatase crystals of considerable size with crystals of other orientations and/or compositions. The percentage of these anatase (001) facets decreases with annealing T while the others increase as indicated by X-rays and AFM measurements [29]. The LaAlO$_3$ (001) (LAO) system has in-plane lattice parameters close to those of anatase (001). The lower mismatch with LAO reduces the crystal quality when the film thickness increases, while in STO it improves due to the relief of the lattice strain [30].

The possibility of employing other techniques such as sol-gel or atomic layer deposition over ordered substrates would be possible. There are two well-known sol-gel procedures to prepare anatase films by dip- or spin-coating. One involves the preparation of an amorphous film that is converted to anatase after a thermal process [31]. The second is based on the early preparation of a sol of anatase nanoparticles, avoiding a thermal process for the crystallization from amorphous titania [32]. In both cases, surfactants can be utilized for the control of the porosity of the resulting film. However, these methods still need more efforts to develop a process for the growth of highly oriented anatase surfaces.

The experimental work has involved the optimization of the synthesis conditions of anatase (001) films grown on SrTiO$_3$ (001) substrates with techniques that allow their growth with a preferential orientation such as PLD and MBE. During this process, some similarities were discovered between both methods that, despite being different, share those steps related to the preparation of the surface substrate and subsequent annealing treatment of the grown films.

Consequently, the main objective presented in this work is to reach flat anatase (001) surfaces with large terrace sizes (larger than 600 nm) synthesized by PLD and MBE. To achieve such a goal, the best growth conditions, temperature (T) and oxygen pressure P(O$_2$) of anatase (001) films on SrTiO$_3$ (001) substrates were determined. Derived from this work, a correlation is established between both synthesis procedures. Furthermore, the evolution of different structural and morphological properties related to the films have been investigated with film growth temperature, such us: 1) grain size and strain (represented by microstrain and the change of in- and out-of-plane lattice parameters), 2) the effect of annealing temperature, and 3) the influence of the oxygen content in the film.

## 2.- EXPERIMENT DETAILS

The crystalline structure and crystallographic orientation of the titania thin films were characterized by X-ray diffraction (XRD) (Malvern PANalytical X'pert Pro MRD) using Cu K$\alpha$ radiation ($\lambda$=1.540 Å). The surface morphology was studied by atomic force microscopy (AFM) (Keysight 5100).

*2.1. SrTiO$_3$ (001) substrate preparation:*

SrTiO$_3$ (001) substrates from Crystec GmbH (miscut < 0.3°, dimensions: 5x5x0.5 mm) were used for TiO$_2$ film growth. The procedure to obtain atomically flat Ti-STO terminated surfaces involves annealing at 1000°C for 1 hour in air and further sonication in deionized water etching for 15 minutes [33]. After this treatment, AFM measurements show Ti-terminated flat terraces with step heights between them of one unit cell, i.e. 3.9 Å. These substrates were placed in their respective PLD/MBE chambers and before starting the growth process they were cleaned to remove residual carbon from their surface due to exposure to air by annealing them at 850 °C at $10^{-6}$ mbar oxygen pressure during 20 minutes.

*2.2. PLD growth:*

A commercially available $TiO_2$ target (diameter: 5cm, purity: 99.99%, Mateck) was used to prepare the titania thin films by PLD. $TiO_2$ films of about 50 nm thick were grown using a 248 nm KrF excimer laser with a repetition rate of 2 Hz and an energy density of ~1.8 $J/cm^2$. The number of laser pulses necessary to grow a 50 nm anatase film is of 1500. The growing process takes 12.5 minutes at a speed rate of 0.033 nm/pulse.

Substrates were introduced in a non UHV-PLD chamber until static vacuum pressures close to $1.33 \cdot 10^{-8}$ mbar were achieved. Then STO substrates were annealed at a $P(O_2)$ of 0.11 mbar up to the $TiO_2$ deposition temperature to guarantee carbon removal. The films were grown in a range of temperatures between 600 and 925°C at identical $P(O_2) = 0.11$ mbar. After deposition, films were cooled down at 7° C/min at $P(O_2) = 13.33$ mbar. The evolution of the film growth was monitored by in-situ reflectivity high energy electron diffraction (RHEED). Ex-situ XRD (θ-2θ) and AFM measurements were performed to study the structure and morphology of the films.

*2.3. MBE growth:*

$TiO_2$ films in the nanometric range were grown at the Néel Institute by Molecular Beam Epitaxy (MBE) using a UHV set-up composed by a preparation chamber coupled to an analysis chamber. The latter is equipped with a commercial scanning tunneling microscope (Omicron VT STM/AFM), a low-energy electron diffractometer (LEED) and an Auger electron spectrometer (AES). The substrate temperature was monitored using an infrared pyrometer (IMPAC Infrared GmbH IPE140). Cleanliness was checked by AES. Metallic titanium was evaporated (Titanium pellets 3x3 mm 99.995%; Mateck) by an electron beam evaporator (EFM 4 Omicron), with the chamber backfilled with molecular oxygen at a partial pressure of $10^{-6}$ mbar at 600°C. The titanium flux was previously calibrated by using a quartz crystal microbalance. The deposition rate was about 0.012 Å/s. The deposition of 8 nm of titanium is equivalent to 16 nm of $TiO_2$ anatase. The same relation is maintained for the other thicknesses.

The STO substrates were treated as described for the PLD samples, but were additionally cleaned by two successive cycles of $Ar^+$ ion sputtering (0.6 keV, $P(Ar) = 5*10^{-6}$ mbar, t=5min and t=10 min) and annealing at 800°C under a $P(O_2)$ of $10^{-6}$ mbar for 20 and 40 min, respectively. This step could be replaced by annealing at 850°C at an oxygen pressure of $10^{-6}$ mbar for 20 min. The cleanliness was checked by AES. $TiO_2$ films of 2, 4, 8 and 16 nm thickness were prepared by depositing Ti at 600°C as mentioned above and characterized in situ by Auger and low energy electron diffraction (LEED) techniques. Subsequently, all samples were annealed ex situ in a cylindrical ceramic furnace at 800°C in 1 bar oxygen for 6 hours and reintroduced into the UHV chamber. Their surfaces were refreshed by performing two successive cycles of $Ar^+$ ion sputtering (0.6 keV, $P(Ar) = 5*10^{-6}$ mbar, t=5min) and annealing at 700°C under a $P(O_2)$ of $10^{-6}$ mbar for 30 minutes. The resulting films were re-analyzed with the same UHV techniques, but only the one with the best LEED diffractogram was inspected by STM.

**3.- RESULTS**

The synthesis of anatase films by PLD has been already described in section 2.2. Willfully, the process developed for the preparation of PLD anatase films (involving substrate preparation and film annealing steps) was also applied to the growth of $TiO_2$ films by MBE. This allows to check the similarities/capacities between the two synthesis procedures and, eventually, to unify the growth conditions of the anatase films or their respective preliminary and subsequent temperature treatments. The last part of each section focuses on the effect of annealing on the grown films.

*3.1. Anatase growth mode*

The growth morphology of the anatase film on STO(001) by PLD has been well described by Radovic et al. [34] and would involve 3 stages during the deposition process: pseudomorphic STO growth, intermediate stage where $TiO_2$ initially grows forming islands and anatase formation (Supplementary information, section II).

Figure 1 shows a schematic view of the proposed structural growth morphology of the anatase (001) film on $SrTiO_3$ (001). The left side of the Figure 1 simulate several processes that take place during growth: Sr-diffusion to the surface, islands formation and step flow growth. In the initial stage of growth, Ti reaches the terrace steps and condenses at the edges during a step-flow growth mode. The intermediate stage is compatible with a Volmer-Weber growth mode in which the interaction between the incoming surface atoms is small, giving rise to the formation of islands. During the initial stage of formation of islands, $TiO_2$ is proposed to grow at the interface between the islands and the upper STO layers by forming misfit dislocations to reduce the mismatch between the in-plane cell parameters of anatase with those of the STO, as observed in anatase (100) films grown on STO(001) [35]. The right image in Figure 1 shows the process in a projected 2D X-Z plane involving 10 STO cells (X direction) and 5 anatase cells along the Z direction, however, the distortions are noticeably reduced after 2 or 3 cells are grown. In this initial slab of 10 STO cells, the reduction of the mismatch between film and STO is 1/3 of the total 3% mismatch between them. The 3rd stage involves growth through the flow of crystal steps where the gaps between adjacent islands are filled [34].

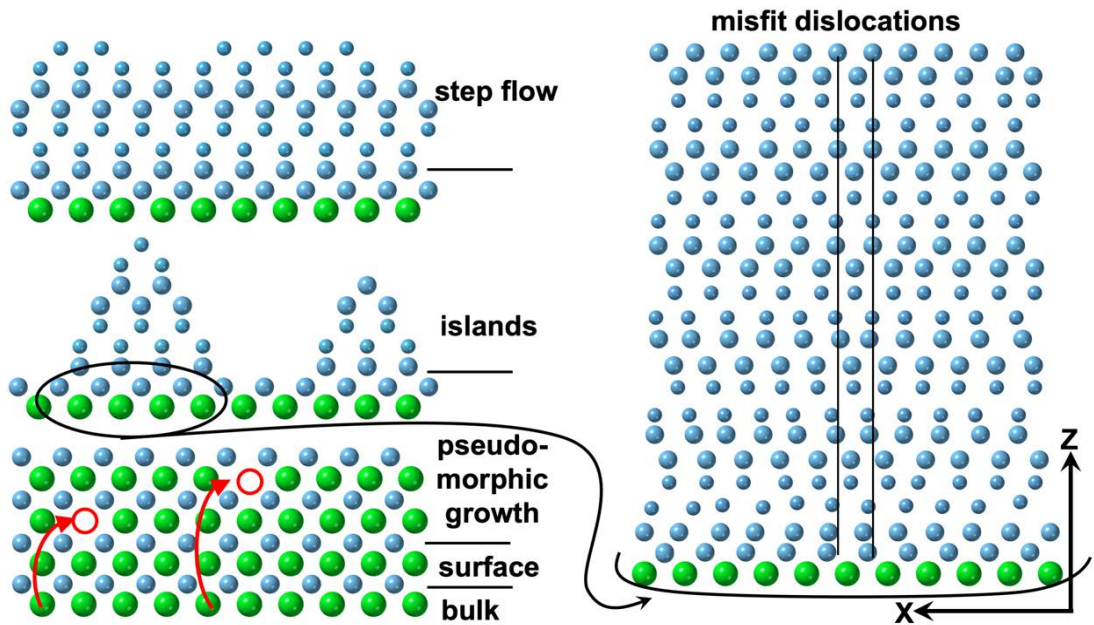

*Figure 1. Schematic diagram of $TiO_2$ anatase (001) growth. Green, blue spheres correspond to Sr and Ti atoms, respectively. The largest/smallest blue spheres correspond to atoms placed in the cell with Y=0 or Y=1/2 coordinates, respectively. The red empty circles correspond to Sr diffusion from bulk to the surface. The images on the left side show an orientation (X, Z) identical to that on the right.*

### 3.2. XRD characterization of PLD films

The growth temperatures of the different PLD anatase films are listed in Table 1. All films have similar thicknesses (about 50 nm) due to the identical number of pulses used during the PLD growth. The structure and surface morphology of sample H was studied after each of the 4 consecutive annealing treatments (6 hours in air each) at 600ºC, 700ºC, 800ºC and 900ºC carried out in a cylindrical ceramic furnace.

| TiO$_2$ film | A | B | C | D | E | F | G | H |
|---|---|---|---|---|---|---|---|---|
| T(ºC) | 600 | 650 | 700 | 750 | 800 | 875 | 925 | 600 |

*Table 1: List of the TiO$_2$ thin films on STO substrates, grown by PLD, with corresponding growth temperatures. Film H was annealed 4 times at 4 different temperatures. Film G was not annealed.*

The analysis of the inner structure of the anatase (001) films was carried out with standard XRD laboratory techniques, i.e. (θ-2θ) and reflectivity measurements. One of the diffractograms, measured for the samples listed in Table 1, shows the indexed peaks relative to sample A (Figure 2a) corresponding to the STO (S00L) and anatase (A00L) reflections ($2\theta_{1,2}$ = 37.6º and 80.5º). The diffractogram shows the specular reflections of the cubic STO cell and those of the tetragonal anatase cell. Their lattice vectors along the (X,Y,Z) directions are parallel to each other, i.e., STO ($a_{STO}$, $b_{STO}$, $c_{STO}$) where $a_{STO} = b_{STO} = c_{STO}$ = 3.905 Å and for anatase film ($a_{film}$, $b_{film}$, $c_{film}$) with $a_{film} = b_{film}$ and $c_{film}$ values listed in table TS1 at the corresponding preparation temperature (T). Figures 2b and 2c show the (004) and (008) anatase peaks before (black) and after (red) annealing at 900ºC during 6 hours in air using a ceramic oven. The anatase peaks move towards smaller 2θ-values which is equivalent to an increase of the *c*-lattice parameter of the film. These 2θ-shifts are also observed in the same anatase films grown on substrates with lower lattice mismatch as a function of annealing temperature [36]. As expected, the peak shift is much smaller for sample F than for sample A due to the proximity of the growth and the annealing temperature, confirming the T dependence of the peak shift.

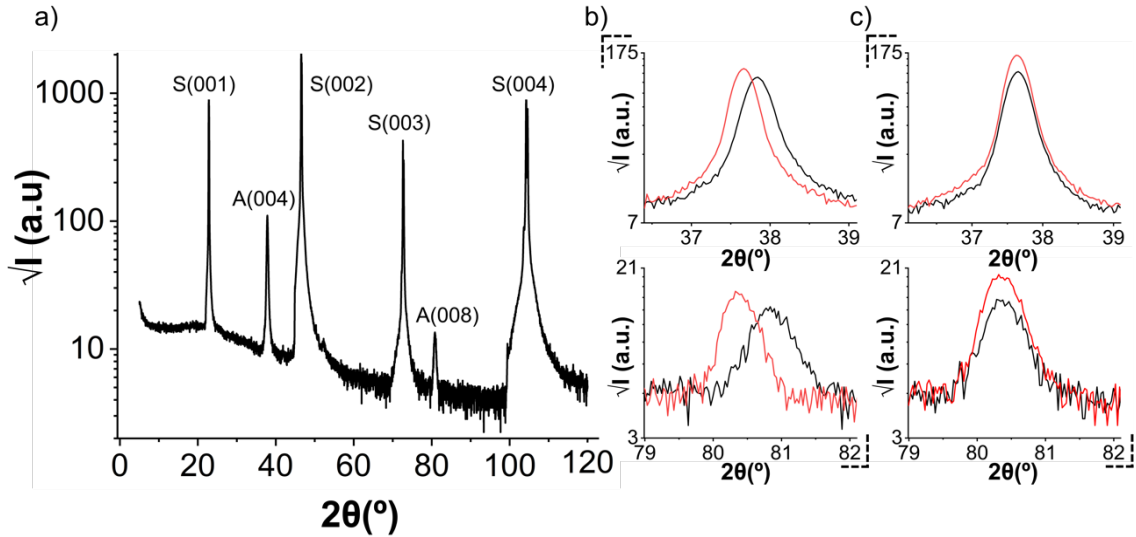

*Figure 2: a) X-ray diffractogram of sample A grown at 600ºC and zoom into the anatase peaks before (black) and after (red) annealing. b) Sample A (600ºC and 900ºC). c) Sample F (875ºC and 900ºC).*

The $c$-lattice parameters ($c_{film}$) of the freshly grown anatase (001) films {from 9.5014 to 9.5464 ± 0.0005 Å in the range 600 – 925 ºC} were measured at room temperature (RT) and normalized to the $c$-bulk parameter ($c_{bulk}$) of the tetragonal anatase crystal ($c_{norm} = c_{film} / c_{bulk}$), ($a_{bulk} = 3.7845$ Å, $c_{bulk} = 9.5143$ Å [37]). Figure 3 shows the values of the normalized $c$-lattice parameters ($c_{norm}$) as function of the growth temperature (GT) (upper black dots curve). *The magnitude of a shift, or equivalently the increase of the c-lattice parameter, depends on the final T applied to the film: GT and annealing treatment.* A more detailed analysis on this point is given in the discussion section. The same Figure 3 shows the evolution of the particle size obtained from the AFM image analysis with the GT (red curve) so, this size defines the lateral dimensions of the islands that form the film. The particle size is not related to the crystal size, as particles can be formed by several crystals. Particle size, grain and island could be considered equivalent, so these terms will be used interchangeably throughout the document.

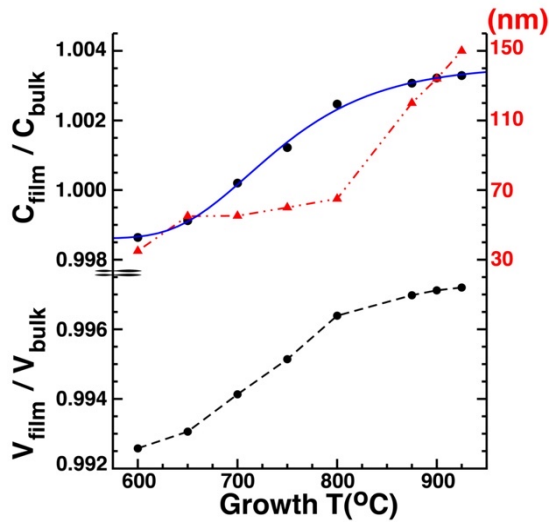

*Figure 3: Evolution of the particle size of the anatase (001) films vs. growth T (red curve); normalized $c_{film}$ dependence on T (black dots); fit of the experimental points to a Gumbel distribution (blue line) [38].*

The $a_{film}$ values of the samples typed in Table 1 were obtained by performing X-rays-RSM (Reciprocal Space Mappings) measurements around the (103) reflection. These values are listed in Table TS1 and indicate constant biaxial compression over the explored T-range. The evolution of the anatase cell volume with T is a relevant magnitude to understand the properties of these films related with their inner structure and stoichiometry. The evolution of the normalized film cell volume ($V_{norm} = V_{film} / V_{bulk}$) is shown in Figure 3 as a dashed black line crossing the experimental black dots. The volume values of the film were calculated using the average in-plane lattice parameter $<a_{film}> = 3.773_3$ Å obtained from the values listed in Table TS1 and those of the $c_{film}$ given in the same figure, i.e. $Vf_{ilm} = <a_{film}>^2 * c_{film}$. Note that the $V_{norm}$ values are less than one indicating that the anatase film cell is under compression in this T-range.

The compressive behaviour of the $a_{film}$ lattice parameter has also been observed in similar thin films [30]. Furthermore, the measurement of similar $a_{film}$ values corresponding to two samples with identical GT (875ºC) using synchrotron diffraction techniques also shown smaller values than those of the bulk, i.e. 3.7784 +/- 0.0005Å for PLD and 3.7811+/- 0.0005Å for MBE (8 nm film). The alignment between the substrate and the film lattice vectors was also confirmed from this experiment, i.e. the (H,K) values of the anatase (10) reflection expressed in STO basis are (1.0335, -0.0008).

The thickness of the films listed in Table 2 were obtained from reflectivity measurements. Figure 4 shows the X-ray diffractogram measured for sample A after its preparation by PLD.

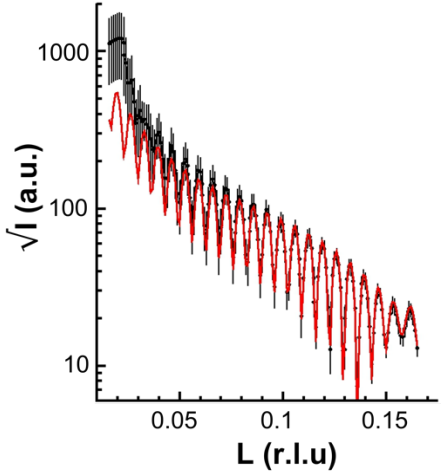

*Figure 4. Experimental reflectivity curve (black) for the anatase (001) film (sample A) in units of the Miller index L corresponding to the STO cell (r.l.u. = reciprocal lattice units). The red curve shows the best fit.*

The data in Figure 4 is expressed in units of L calculated using the $c_{STO}$=3.905 Å lattice parameter. In this reciprocal space notation, the film thickness was calculated using the expression:

$$t = \sum_{n=1}^{m} \Delta\, c_{STO}\, n\, [TiO_2]_{STO} \qquad (1)$$

where $n$ is the number of $TiO_2$ cells forming the film, $[TiO_2]$ includes the atomic coordinates of the anatase (001) unit cell expressed in units of the STO, $\Delta$ is an offset of the anatase diffractogram, or equivalently, a normalized parameter that modifies the $c_{STO}$ lattice value used to define the $TiO_2$ cell to correct for the diffractogram shift along L. In this base, the anatase $TiO_2$ $c_{bulk}$ value is 2.4364 times greater than that of the $c_{STO}$. The resulting film thickness is approximately 54.1 nm corresponding to $n$ = 57 anatase cells with $c_{film} = c_{bulk} \Delta$ = 9.5143 * 0.9985 Å or 57 $TiO_2$ cells expressed in STO units: 57 * (2.4364 * 3.905)*0.9985 Å = 54.1 nm.

The film thicknesses and crystallite (or crystal) sizes of the samples listed in Table 2 were calculated from the θ-2θ XRD diffractograms using Snell's law, Debye-Sherrer and Williamson-Hall (WH) formalisms, respectively [39-41]. The table also includes the microstrain ($\varepsilon_{mic}$) values for each of the film growth temperatures, the particle sizes measured by AFM and some terrace sizes after annealing at 900°C (G: 925ºC) for 6 or 12 hours. Note that the crystal size and strain values calculated at each temperature via the WH formalism only use the two anatase reflections contained in the measured θ-2θ X-ray region. The experimental errors reported in the last two columns of Table 2, referring to island and terrace sizes, were obtained by calculating the root mean square deviation of a sampling including 5 horizontal line profiles (5 microns long) from AFM images.

The θ-2θ diffraction patterns of anatase films are characterized by the presence of peak broadening and by shifts of these peaks from the ideal (or bulk) positions. The experimental profile of the diffraction peaks results from the convolution of the intrinsic profile with the instrumental contribution. Therefore, once the instrumental contribution is removed, the remaining contribution to peak broadening is caused solely by the intrinsic properties of the material, *e.g.* the presence of uni- or multimodal distributions of small crystallites (coherent diffraction domains) or structural imperfections (crystalline distortions, dislocations, stacking defects,). In the case of heteroepitaxial films such as the anatase of this work, the peak widths only reflect the projection of the magnitudes of these intrinsic properties along the surface normal direction. In this work, all causes contributing to the additional peak broadening (with respect to the proper width of the film caused by its crystalline size) will be interpreted in terms of microstrain ($\varepsilon_{mic}$) [42] which can be understood as the presence in the film of a lattice constant distribution due to the crystalline defects. X-ray broadening can be used to investigate the microstrain or the defect distribution in the film [43-45]. Conversely, when strain is uniform throughout the lattice, lattice deformation manifests itself as a shift of the diffraction peaks ($\varepsilon_{Lat}$) [46,47].

The main difference between the respective θ-2θ diffraction patterns of epitaxial films and powder samples concerns the structural information they contain. In epitaxial films, the θ-2θ patterns provide access to structural information along one direction as, for example in the present case, where only reflections (00L) are available. In contrast, indexed powder diffraction patterns allow the study of crystal size and strain in all directions.

Some of these defects commented above are reduced during annealing, as indicated by the decrease in microstrain with annealing temperature performed for a single sample, as shown in Table 3.

| $TiO_2$ film | Thickness (±1 nm) | D (Sherrer) (± 6 nm) | D (WH) (± 6 nm) | $\varepsilon_{mic}$ (WH) | Grain size AFM (nm) | Terrace size (nm) |
|---|---|---|---|---|---|---|
| A | 52 / 54[*] | 32 | 39 | $0.00127_{10}$ | $35_7$ | $120_{25}$ |
| B | 44 | 25 | 41 | $0.00172_{10}$ | $55_{10}$ | $300_{50}$[*] |
| C | 49 | 34 | 40 | $0.00105_{10}$ | $57_{10}$ | |
| D | 50 | 29 | 50 | $0.00146_{10}$ | $60_{10}$ | $400_{70}$[*] |
| E | 44 | 26 | 39 | $0.00135_{10}$ | $64_{10}$ | |
| F | 51 | 32 | 49 | $0.00161_{10}$ | $120_{25}$ | |
| G | 54 | 28 (<30>) | 45 (<40>) | $0.00145_{10}$ | $140_{27}$ | $650_{100}$[**] |

*Table 2. Content: thickness values calculated using the Snell's law, the asterisk indicates the value obtained from expression (1), and crystallite sizes (D) obtained using the Debye-Sherrer expression or Williamson-Hall method for the as-prepared samples listed. Their averages are indicated in the last row. The errors were obtained from the discrepancies between the used methods: Snell/diffraction and Sherrer/Williamson applied to their respective distributions. The microstrain ($\varepsilon_{mic}$) in fresh films does not follow a clear trend with T. The last column shows the evolution of the terrace sizes measured by AFM after 6 or 12[*] hours annealing at 900ºC. [**]Sample G was annealed at 925ºC during 12h.*

A detailed XRD analysis of the evolution of the widths (β) and intensities subjected to various annealing treatments at different temperatures (6 hours each) was performed for sample H. Line broadening analysis of anatase film peaks performed using integral breadth (IB) methods [48,49] in combination with the Williamson-Hall method [41] offers a valuable tool for a preliminary assessment of the IB-related effects of crystallite size and lattice strain. This formalism has been applied to the anatase $2\theta_{1,2}$ peaks as shown in Table 3. The estimated errors associated to these angles were obtained from their sensitivity to induce changes in strain. Similar procedure was followed for the estimation of the strain errors. The normalized $c_{film}$ lattice parameters are obtained from the $2\theta_1$-positions of the table.

| T (º) [a,b] | $\beta_1$ / $2\theta_1$ (º) | $\beta_2$ / $2\theta_2$ (º) | $\varepsilon_{mic}$ | D (Å) | $c_{film}/c_{bulk}$ |
|---|---|---|---|---|---|
| 600 [a] | 0.3520 / 37.851$_1$ | 0.5823 / 80.852$_1$ | 0.00127$_{10}$ | 392$_{10}$ | 0.9985$_1$ |
| 600 [b] | 0.3513 / 37.831$_1$ | 0.5817 / 80.790$_1$ | 0.00127$_{10}$ | 394$_{10}$ | 0.9999$_1$ |
| 700 [b] | 0.3223 / 37.788$_1$ | 0.5177 / 80.681$_1$ | 0.00096$_{10}$ | 403$_{10}$ | 1.0001$_1$ |
| 800 [b] | 0.3147 / 37.699$_1$ | 0.4809 / 80.480$_1$ | 0.00065$_{10}$ | 376$_{10}$ | 1.0024$_1$ |
| 900 [b] | 0.3093 / 37.671$_1$ | 0.4723 / 80.385$_1$ | 0.00062$_{10}$ | 382$_{10}$ | 1.0031$_1$ |

*Table 3. Integral breadths measured on sample H versus temperature. a and b super-indexes indicate the temperature of the 'as-prepared' or the annealed sample. The β-widths have been calculated using the expression β = Area /I$_{max}$, from the area and maximum intensity of the peak (in units of detector counts). The values of the microstrain (ε$_{mic}$) and the crystallite size (D) were obtained from the Williamson-Hall method. The microstrain reduces with T as already reported [50,51] while c$_{film}$ increases (last right column).*

The maximum GT explored, as indicated in Table 1, was 925ºC. However, the effect of high temperature annealing (T > 950ºC) was also investigated with the intention of establishing a maximum annealing temperature value before the anatase to rutile phase transition appears. This transformation starts to appear at 950ºC showing weak contributions from the rutile and brookite phases which add to the anatase diffractogram, Figure S4.

The X-ray diffractogram of sample E after annealing it at 1000ºC, Figure 5, shows some reflections coming from the 3 polymorphs of TiO$_2$. According to its indexation, the rutile phase shows the (110) reflection with respect to the STO substrate while brookite shows reflections of grains with orientations (001) and (100) although this phase looks residual when comparing its intensity with those of the anatase and rutile. The figure is expressed in units of the L-Miller index, as it is intended to reflect that the displacements detected in the anatase peaks are not detected in the STO peaks, as their positions coincide with integer L-values.

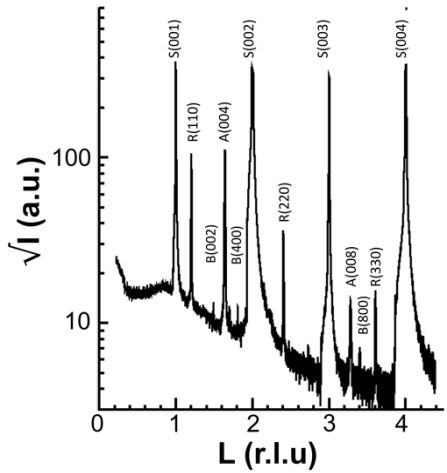

*Figure 5. XRD diffractogram of sample E annealed at 1000º C. Main TiO$_2$ phases are rutile (R) and anatase (A). Weaker peaks belong to the brookite phase (B).*

The first anatase peak in Figure 5 is at $L_{E,1}$=1.642 while the same peak in sample F is found at $L_{F,1}$=1.637. The calculation of these L-values on the anatase base using the expression $c_r = c_{bulk}/c_{STO}$ = 9.5143 / 3.905, gives the values $L_{E,1}' = L_{E,1} * c_r = 4.000_6$ and $L'_{F,1} = L_{F,1} * c_r = 3.988_5$.

The influence of P(O$_2$) during the growth process was also verified. Samples prepared at a lower P(O$_2$): 0.013 mbar, show similar x-ray diffractograms and average grain sizes as those prepared at higher pressures. The main difference between the samples grown at low and high P(O$_2$) is observed in their respective topographies, which is more homogeneous when pressure is fixed at 0.11mbar. A further increase in P(O$_2$) to 0.153 mbar does not improve the surface morphology and the PLD pumping system becomes unstable.

The P(O$_2$) can also modify the $c_{film}$ magnitude of the prepared films. Figure S5 shows the effects of oxygen pressure during the growth process in $c_{film}$ for two different cases. The oxygen deficit, that is, P(O$_2$): 0.013 mbar, increases the normalized $c_{film}$ value up to 1.01, much higher than those shown in Figure 3. Oxygen pressures greater than 0.11 mbar do not cause additional changes in the $c_{film}$ (Figure S5).

### 3.3. AFM characterization of PLD films

The surface morphology of the PLD films, i.e., the evolution of the lateral size of the TiO$_2$ particles with deposition and T-annealing was investigated by AFM as detailed in Figure 3 (red curve). The size of the islands observed on the surface of the anatase films measured by AFM is highly dependent on GT, as shown in Figure 3 (red curve). Film A prepared at 600ºC shows the smallest grain size (~35nm) while film G grown at the highest T (925ºC) exhibits the largest island size (~150 nm), see AFM images in Figure 6. In the temperature range between 600-800ºC the increase in grain size is small (only 25 nm), however, above 800ºC the size increases markedly.

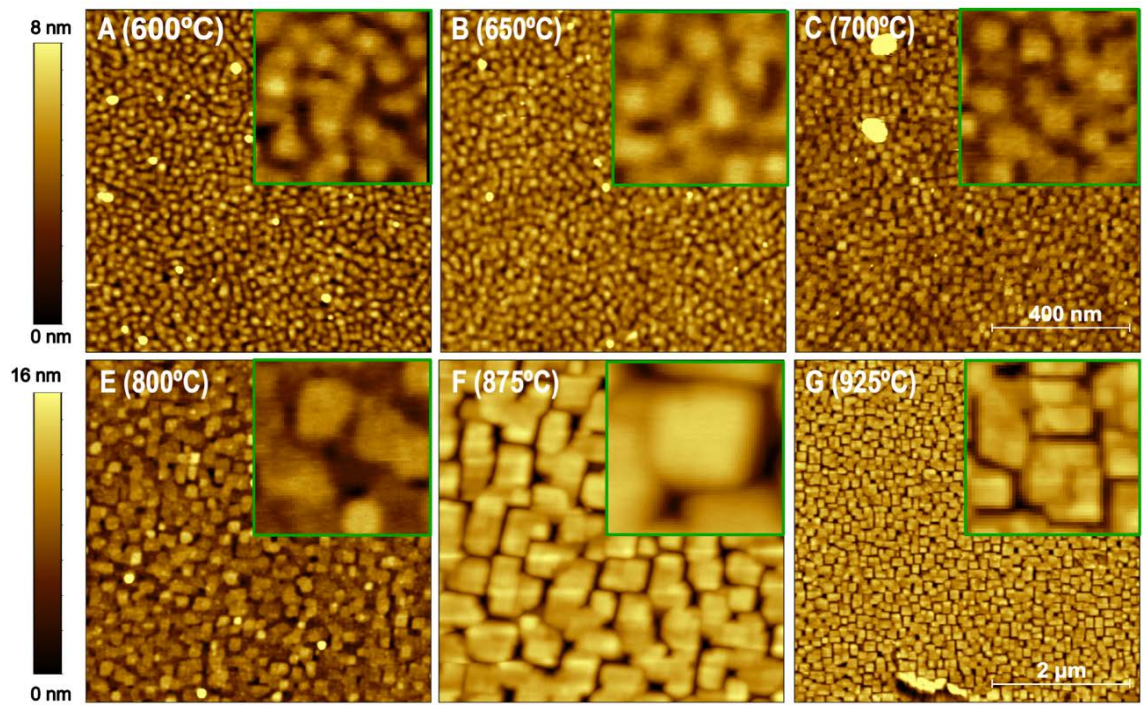

*Figure 6. AFM images of TiO$_2$ PLD-fresh prepared films at different growth temperatures for the indicated samples. The sizes of images [(A to F), and (G)] are [(1x1 µm; zoom: 250x250 nm) and (5x5 µm; zoom: 1x1 µm)], respectively.*

#### 3.3.1. Effect of heat treatment on PLD films

A relevant data obtained from the $c_{film}$ evolution with T (Figure 3) is the indication of the temperature range (800-900ºC) in which the annealing process efficiently transforms the degree of crystallinity of both the film and its surface. Post-growth annealing alters the surface grain morphology of the films into larger flat terraces.

The overall effect of T-treatments after deposition has been described using sample H. This film initially presents a surface distribution of small grains with an average size of 35 nm which transform into larger flat terraces of about 120 nm after several annealing treatments. This evolution can be observed in Figure 7 and reveals the possibility to obtain large flat ordered terraces (around 120 nm) quite independent of the initial grain size obtained at their corresponding GT. Successive long annealing steps (6 hours each) at progressively increasing temperatures in ambient air conditions have the effect of smoothing the film surface by replacing the surface grain morphology by a flatter plate-like one. Figure 7a shows the surface grain morphology of anatase from the initial grown film. The circular-shaped grain morphology changes to a square-shaped morphology after annealing the film at 700ºC (Figure 7b). At 800ºC, the grains coalesce into plates (Figure 7c). Maximum coalescence is achieved at 900ºC giving the largest possible terrace sizes (Figure 7d) in this series of measurements.

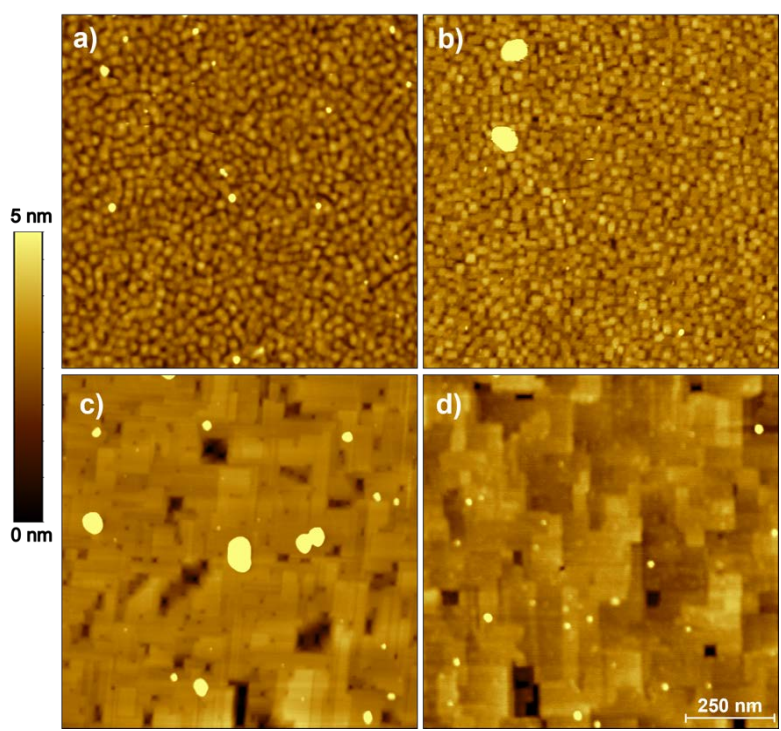

*Figure 7. AFM images of sample H: (a) 'as-prepared' (600ºC), and annealed (b) 700ºC, (c) 800ºC, and (d) 900ºC.*

The terrace sizes shown in Figure 7d are similar to those obtained by MBE by Du et al [28]. In both cases the T and annealing time (900 °C, 6 hours) used in PLD were equivalent to the T and deposition time (750 °C, 6 hours) used in MBE. Both procedures show similar surfaces where the islands coalesce to form flat terraces of approximately 100 nm. However, the internal structure, although similar, should not be stoichiometric $Ti/O_2$ in the MBE case, as demonstrated in Figure S5. The oxygen pressure used during the growth process was $10^{-8}$ mbar, below the critical value of 0.1 mbar shown in figure S5 for PLD, and below the $10^{-6}$ mbar (section 2.3) used to grow already oxygen-deficient thin films by MBE. In any case, it is shown that similar film quality can be obtained using PLD and MBE in a two-step process: first growing the film to a given T (in oxygen-saturated atmosphere if possible at least for PLD) followed by annealing at ambient pressure at temperatures above that of coalescence (> 800 °C). The MBE growth process as ref. [28] represents the conventional option where operating conditions, $O_2$ pressures and T, must be compatible with UHV.

The effect of larger annealing times can be seen on Figure S6 (sample D). Twelve hours of annealing at 900º C generates terraces with sizes of approximately 400 nm.

A similar 12-hour annealing of sample G at temperatures of 925°C increases the flat terraces to 650 nm, Figure 8. This figure shows the AFM image of sample G grown at 925°C (left image) and subsequently annealed in air for 12 hours (middle image: 3h temperature raise from room temperature to 925ºC, 6h constant T at 925ºC and 3h drop to RT). The film surface shows the evolution from the typical distribution of square geometry islands formed during high temperature growth (Figure 6G) to long flat terraces after annealing in air. Comparing this annealing with those performed in sample D (Figure S6) the size of flat terraces increases from 400 nm to 650 nm.

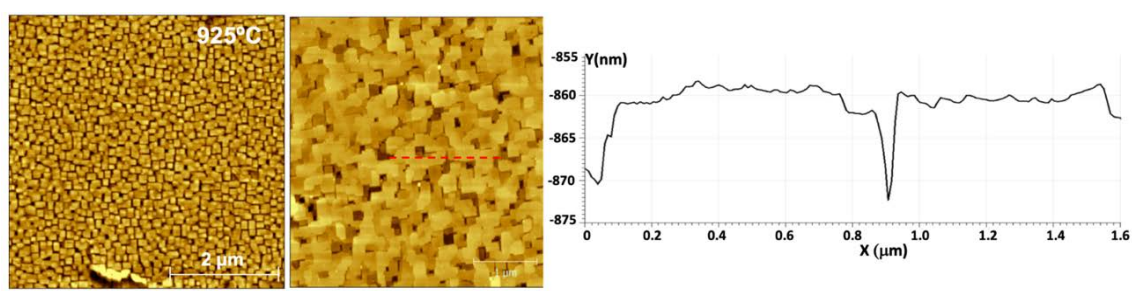

*Figure 8. AFM images of sample G after PLD preparation (left) and annealed at 925ºC during 12 hours (middle). Line profile along the surface of the aneealed sample (right). The average surface terrace size is about 600 nm.*

*3.4. LEED/STM characterization of MBE films*

$TiO_2$ films of different thicknesses (2, 4, 8, and 16 nm) were grown by MBE under identical T and oxygen pressure conditions. Ti was evaporated by keeping the STO substrate at 650°C and the chamber at $10^{-6}$ mbar $O_2$ pressure. At this stage, all prepared MBE films shown a diffuse LEED diffractogram with broad spots evidencing the formation of a poorly ordered surface structure. Figure 9a shows the LEED diffractogram of an 8 nm thick film. The samples obtained by MBE were subsequently annealed ex-situ at 850 °C in oxygen (ambient pressure) for 6 hours. The films were returned to the UHV chamber and were sputtered/annealed ($Ar^+$ $10^{-6}$ mbar, energy 500 eV, 2 minutes) / (850 ºC at $10^{-6}$ mbar of $O_2$ pressure). After this process, the LEED image of the 8 nm thick sample shows a sharp pattern with two domains of the well-known (4x1) reconstruction routinely observed in anatase (001) single crystals [52] (Figure 9b). Only in the thinnest deposit (2 nm) no LEED pattern was observed. To compare the surface structures of MBE and PLD films, a PLD film grown at 875°C was introduced into the UHV system and studied by LEED. This shows a diffuse pattern and faint spots (1x1) (Figure 9c); however, after a soft sputtering/annealing cycle under conditions similar to those used to restore the surface of the MBE samples, a comparable LEED pattern (4x1) was also obtained (Figure 9d). This procedure shows that PLD and MBE techniques can be used to synthesize anatase (001) films with very similar surface structure.

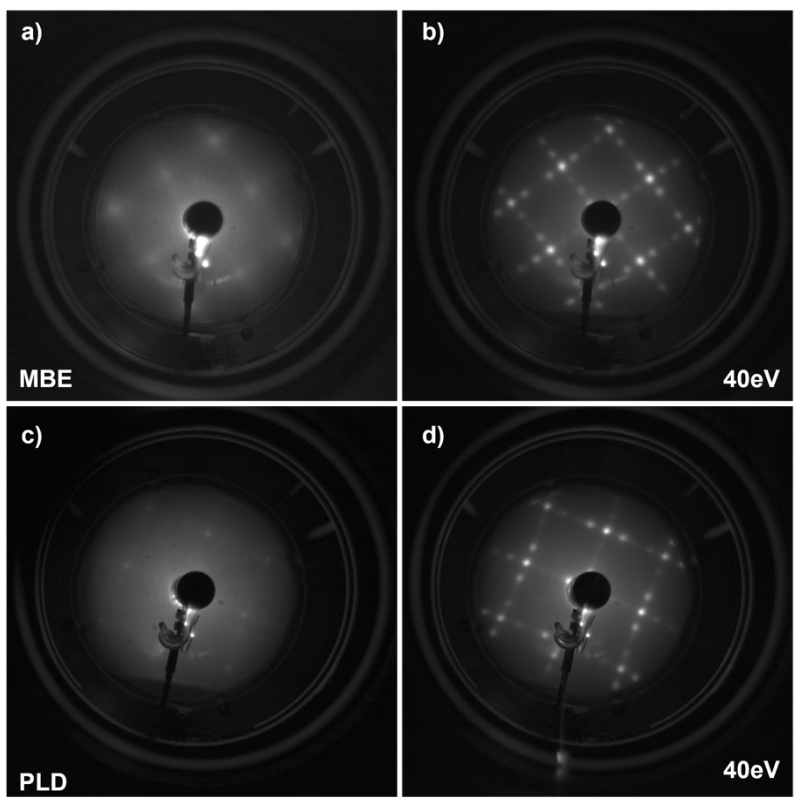

*Figure 9: LEED diffractogram of an 8 nm thick anatase film prepared by MBE (top): (a) sample as deposited and (b) sample after annealing ex-situ at 850ºC with. sputtering/annealing treatment. LEED diffractogram of a 55 nm thick anatase film prepared by PLD (bottom): (c) sample as deposited and (d) sample after sputtering and annealing at 850ºC in $O_2$.*

To better understand the reconstruction (4x1), the atomic structure and surface morphology of anatase films were investigated using the STM technique. Figure 10a shows an AFM image of the surface of an 8 nm thick anatase film. The macroscopic AFM image of the PLD-grown samples (Figure 7) shows flat terraces of about 100 nm, probably formed by two smaller grains than those observed in the STM images (Figures 10b-c). These images show well-defined regions of the anatase surface showing parallel rows along the main crystallographic directions of the substrate surface with average gaps between them of approximately 1.6 nm, about four times the unit cell distance (1x1) ~15.12 Å. The images also show smaller crystal sizes of the anatase surface than those of the PLD samples, probably due to the difference in thickness between them. Indeed, a certain amount of material (thickness) is needed to fill the spaces between the grains formed during growth. These "square" shaped grains, easily identifiable on the AFM image of PLD films, accumulate holes or spaces between them that decrease in number after high-temperature annealing, where the spaces between grains are substantially filled during their coalescence (Figures S6 and 10). The topography of the samples also indicates a correlation between the amount of material deposited with the maximum achievable domain size [28].

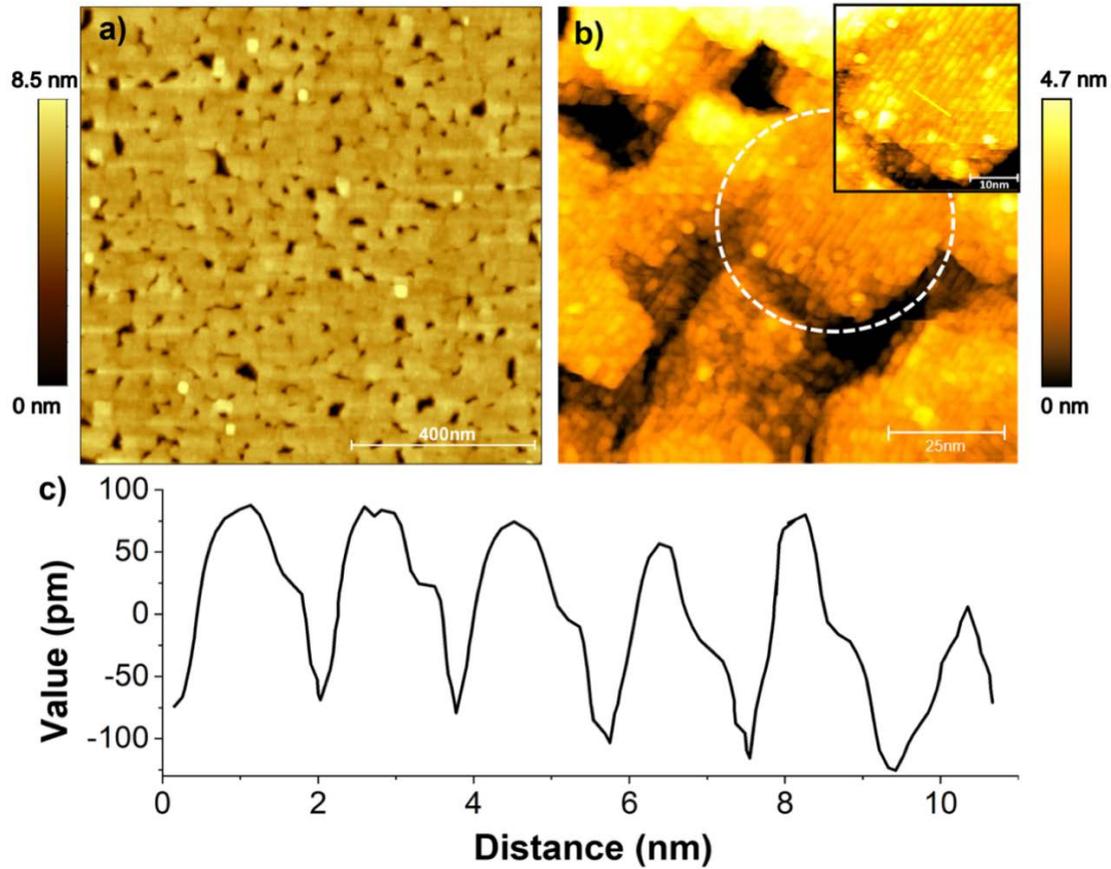

*Figure 10. (a) 5x5 µm² AFM image of a 8 nm thick film (b) STM images of a 8 nm thick anatase (001) film measured at several resolutions. The upper image corresponds to a zoom of the marked $TiO_2$ grain. At the nanometer scale the typical rows of the (4x1)/(1x4) reconstruction are observed (c) line profile of the yellow line.*

## 4.- DISCUSSION

*4.1 X-ray diffraction analysis*

The X-ray diffractograms in Figure 2 reveal that the anatase peaks: shift towards lower Bragg angles evidencing an increase in $c_{film}$ with T (Figure 2b); the increment between peak shifts is proportional to the difference in T; they reduce their widths (as remarked in Table 3); and increase their intensity. In addition, Table 3 shows the strongest decrease in peak widths in the T-region ≥ 700°C, which is related to an increase in film crystallinity along the [001] direction. Furthermore, upon annealing at high temperatures (Figure 2c) the anatase peaks continue to increase in intensity. However, the major structural rearrangement occurs when temperatures above 800°C trigger the coalescence process, which fuses the crystalline grains that make up the film to form large terraces on the surface, visible by AFM. This lateral structural rearrangement cannot be observed in the θ-2θ diffractograms.

The evolution of the peak shifts of anatase (001) films with T (growth or annealing), as mentioned in section 3.2, is related to a continuous expansion of $c_{film}$. The evolution of the normalized ($c_{norm}$) $c_{film}$-parameter is defined by the upper black dotted curve in Figure 3. Below 700°C, $c_{norm}$ is smaller than one, indicating a compression of the $c_{film}$ parameter of 0.20% at 600°C (compared to the bulk value at RT). Conversely, at higher growth temperatures $c_{film}$ expands by up to 0.33%, i.e. at around 700°C the out-of-plane lattice parameter changes from a compressive to a tensile strain regime. This type of regime change has already been observed in anatase films [50].

The continuous blue curve in Figure 3 represents the calculated $c_{norm}$ values derived by fitting the experimental ones to a cumulative distribution function (CDF) of the Gumbel distribution [38] defined by expression (2)

$$c_{norm}(calc) = p + q \left(e^{-e^{-(x-\mu)/\beta o}}\right) \quad (2)$$

The exponential part, $CDF(x: \mu, \beta eta) = e^{-X(x)}$, is a probability distribution function defined between 0 and 1. Here, $p$ denotes a background baseline that shifts the function to values of $c_{norm}(x)$ close to 1; $q$ is a scaling factor able to reduce the Gumbel distribution in the range $0.990 < c_{norm}(x) < 1.004$; x is the temperature variable. The quantities defining the distribution are the *standard deviation* = $\beta o\, \pi / \sqrt{6}$ with $\beta o = 72.2247_9$ and the *mean* = $\mu + \gamma\, \beta o$ with $\gamma \sim 0.5572$ (Euler constant) and $\mu = 711.0820_9$. $\beta o$ and $\mu$ result from the adjustment to the experimental data (correlation value = 0.9989).

The microstrain values shown in Table 2 do not follow a clear dependence on the GT of the film. The crystallite sizes (Table 2) derived from the WH-method are close to 40 nm, in good agreement with the thickness values obtained from the XRR measurement for the same samples. The crystallite sizes obtained from the Sherrer equation are systematically smaller than those of the WH method, probably because the calculation does not take into account the microstrain of the film. Thus, the WH method gives out-of-plane crystallite sizes, as the reflections do not include the H and K components, and microstrain values that are of the same order of magnitude as those of the c-lattice strain shown in Figure 3. Table 2 also shows the evolution of the terrace size after high temperature annealing at 900 ºC for 6 hours (sample H) or 12 hours (sample G). The effect of prolonged anneals at 900-925ºC is to increase the initial surface size of the islands obtained at the corresponding growth temperature (600ºC < T < 875ºC) to much larger terrace sizes as a consequence of inter-island coalescence.

The effect of annealing-T on strain ($\varepsilon_{mic}$) in anatase films from peak line broadening analysis using IB-methods is shown in Table 3. These values are different, but similar in magnitude, to those of ($\varepsilon_{Lat}$) obtained from the conversion of the evolution of $c_{film}$ with T, from Figure 3 (Table 2) using the expression:

$$\varepsilon_{Lat} = \frac{c_{film} - c_{bulk}}{c_{bulk}} \quad (3)$$

where $c_{film}$ and $c_{bulk}$ are the *c*-lattice parameters of the anatase film and bulk, respectively.

The maximum $c_{film}$ increase over the temperature range is 0.474%. The $c_{film}$ strain varies from -0.0015 (600ºC) < $\varepsilon_{Lat}$ < 0.0034 (925ºC), values which are of the same order as the microstrain values obtained by the WH methods in tables 2 and 3.

Note the differences between the WH microstrain values given in Tables 2 and 3. This presumed discrepancy observed in the different microstrain evolution with temperature calculated using this method is not such since the values in Table 2 are from fresh films grown at different temperatures, while those of Table 3 come from a single sample that has been annealed several times at different temperatures. As mentioned in section 3.2, film reflections contain two different types of deformation contributions, $\varepsilon_{mic}$ and $\varepsilon_{Lat}$, which could be related to each other.

The progressive decrease of the microstrain versus T in Table 3 is due to the annealing effect that reduces the number of defects and microdistortions present in the films, manifesting as a decrease of the peak widths, while the evolution of the WH microstrain values in Table 2 can be considered constant with T because the peak widths oscillate around a mean value, i.e. (<FWHM(004)> = 0.34º ± 0.01º). In this case, similar levels of defects are generated in the freshly prepared films that are independent of temperature due to the rapid growth process associated with the PLD technique [53]. On the other side, the progressive decreasing of the 2θ peak positions in Table 3 is compatible with a progressive expansion of the $c_{film}$ parameter whose normalized values are very similar to those shown in Figure 3 for identical temperatures.

The overall effect of annealing is the reduction of temperature-dependent structural/morphological stress components accumulated throughout the film during its PLD growth. This behavior is quantified in Table 3 as a WH-strain ($\varepsilon_{mic}$) that decreases with annealing temperature. The values of the widths and positions of the reflections collected in this table can be divided into two temperature regions: less (R1) or greater (R2) than 700ºC. The boundary between them marks the change of the strain regime of the c-lattice parameter from compressive to tensile. In the R1 interval, the morphology of the islands changes slightly from a spherical cap (Figure 7a) to square shape (Figure 7b) while the β-widths of the peaks become narrower (Table 3). The square shape of the islands is clearly visible in AFM images measured on films grown at high temperatures (Figure 6: samples F and G). In the R2 region, the square islands merge, restructuring the surface and forming larger terraces that generate a further reduction in structural strain [54,55]. This release of microstructural deformation seems to be related to an increase of the $c_{film}$ strain, as the c-lattice parameter increases (last column of table 3). However, this would not be entirely true in this system, since annealed or grown anatase films at the same temperature show similar levels of $c_{Lat}$ strain, as shown in Table 3 and Figure 3. Due to the lattice mismatch between the substrate and the film, an intrinsic lattice strain appears in the film, which would normally increase with thickness and relax above a certain value [38]. In our 50 nm films, the out-of-plane lattice strain component evolves from compressive to expansive with T, keeping the film volume in the compressive regime (lower dashed black curve in Figure 3) over the whole T range. The decrease of the microstrain with annealing T is not able to reduce the lattice strain generated during growth by the lattice mismatch.

*4.2 Influence of biaxial lattice strain*

The type of strain in the film, compressive/tensile, may be largely due to the difference in lattice parameters between the film and the substrate. If the misfit between both can be kept in the range ±7% fully strained lattice is considered to be possible [56]. In fact, the same system such as $SrRuO_3$ can grow with both types of strain depending on whether it grows in $SrTiO_3$ (001) or $DyScO_3$ (110).

Anatase (001) films grown on STO in the T-range of 600 to 925 °C show in-plane ($a_{film}$) biaxial compressive strain, while out-of-plane ($c_{film}$) compressive strain occurs at temperatures below 700 °C; at higher temperatures $c_{film}$ changes to tensile strain. Biaxial compressive strain could then be the driving force favouring the growth of (001) anatase on STO(001) [57]. This surface possesses the largest surface stress among all rutile/anatase surfaces and occurs along the [100] direction [58]. Consequently, the relative stability of (001) surface can improve under the effect of in-plane compressive strain. Experiments applying external compressive biaxial strain show that the increase in the proportion of the anatase surface (001) is favoured [57]. The origin of this compressive stress could be attributed to the effect of the surface stress inducing compression in the islands before they become firmly attached to the surface. The compressive strain in the islands could be maintained and propagated throughout the growth process [59]. The inter-island stress at their borders could help to its stabilization along the film. Furthermore, it is also reported that c-lattice strain depends on the thickness of the film, showing its compressive character for thin films; this behavior was even detected in the plane for LAO substrates that have smaller mismatch than STO [30]. Moreover, the same system synthesized using sol-gel technique on STO shows a range of temperatures from 700ºC to 1100 ºC where $c_{film}$ shows a progressive and continuous increase in tensile strain [50].

*4.3 Relationship between lattice strain and oxygen vacancies in anatase films*

The values of the cell lattice parameters could delimit or suppress the presence of oxygen vacancies in the film; conversely, their presence could also modify the magnitude of the cell volume. According to this premise, the evolution of the lattice strain associated with the anatase growth process, as a function of T shown in Figure 3, could be assigned to different concentrations of oxygen vacancies during film preparation. In order to know if the prepared films presented in this work are affected by a dependence of the oxygen vacancies with T, or if the lattice values act as a conditioning factor to avoid their presence, the literature on the matter will be revisited and some experimental measurements, already presented in the results and supplementary material sections, will be discussed to clarify this point.

In agreement with the experiments, the diffusion coefficient of oxygen ions is found to substantially increase under tensile biaxial stress and to decrease under compressive biaxial stress [60]. This effect has been observed in various oxide film systems as $CeO_2$ [61] and $SmNiO_3$ on STO and LAO [62]. More specifically, [63] investigated that oxygen defects increase lattice stress and expand the vacuum-annealed ceria lattice. However, with heat treatment these oxygen defects are removed and grain growth occurs.

In this work, the grown films show compressive biaxial strain over the range of temperatures explored, as shown in Table TS1, while the out-of-plane lattice parameters show a change from compressive to tensile strain with increasing temperature. Figure 3 shows the evolution of normalized $c_{film}$ (upper curve with solid black cicles) and $V_{film}$ (lower dashed black curve) against the growth temperature. The film cell volume was calculated considering the average value $<a_{film}>$ = 3.773$_3$ Å obtained from Table TS1, and the values of $c_{film}$ given in Figure 3.

The film cell volume varies from 135.26 Å$^3$ at 600°C to 135.89 Å$^3$ at 925°C, while the bulk cell volume is 136.28 Å$^3$ at RT. The cell volume of the film remains under compression throughout the entire temperature range (Figure 3). This behavior would not agree with a film with a distribution of oxygen vacancies homogeneously distributed throughout the film, since such a distribution would produce an increase in the cell volume of the anatase.

Figure S5 shows the reflection (004) (red curve) of two films grown at 600 °C at two different $P(O_2)$ of 0.013 mbar (red curve) and 0.11 mbar (black curve); this last pressure is the one used to grow the films presented in this work. The oxygen deficient film increases the normalized $c_{film}$ from 0.9982 (0.11 mbar) to 1.010 (0.013 mbar), a much higher value than any of those shown in Figure 3. The $c_{norm}(PO_2 = 0.11 mbar) = 1.010$ generates a $V_{film}$ (= 136.80 Å$^3$) > $V_{bulk}$ which could be related to the presence of oxygen vacancies in the film. The blue and green dashed curves, in Figure S5, correspond to two films grown at 875 °C and $P(O_2)$ of 0.11 mbar and 0.153 mbar, respectively. In this case, both reflections have identical values of 2θ. This experiment proves that the oxygen pressure selected to grow the anatase films is adequate to avoid oxygen vacancies in the films.

Furthermore, the comparison between the $c_{norm}$ values of freshly grown and annealed films at different temperatures, in Figure 3 and Table 3, are very similar. This means that a possible loss of oxygen during the annealing treatments would give higher normalized lattice parameters than freshly grown samples, and this assumption has not been experimentally detected.

These results show that the biaxial compressive strain detected in the films, as well as the behaviour of the compressive strain along the out-of-plane direction at temperatures below 700°C, are not an artefact generated by the presence of oxygen vacancies in the films.

*4.4 PLD versus MBE growth techniques*

Film characterization using diffraction techniques (synchrotron and conventional) show that the [100], [010] and [001] directions of the anatase film are aligned with those of the STO(001) substrate, as mentioned in section 3.2, and the orientation of the terraces follow this alignment.

The SrTiO$_3$ (001) substrates used in this work are TiO$_2$-terminated with a double-height layer between neighbouring steps, with inter-step terrace widths of 250 nm (miscut ~ 0.1° measured from AFM images). The two-step cleaning process, consisting of an annealing followed by a sonic bath in ultrapure water applied to the STO substrates, produces atomically flat Ti-terminated surfaces with a roughness level of less than 0.01 nm. The main advantages of using this process are related to the safety of the chemicals required, as well as the absence of vacuum conditions, which facilitates handling. In addition, samples can be stored in low-pressure chambers for a long time, whose surfaces can be regenerated in the PLD chamber by pre-annealing at moderate temperatures in oxygen, i.e. up to 850°C for one to two hours. However, the use of alternative etching procedures [64] should be explored for the establishment of comparisons between the resulting TiO$_2$ films obtained by PLD and MBE from different substrates prepared with STO.

Figure 6 demonstrates that the PLD growth time, even at high temperatures, cannot be considered an annealing since the morphology of the islands is visible over the entire range of T explored, i.e. up to 925°C. In addition, Figure S6 also shows that island coalescence already takes place at annealing temperatures of 800°C, so any PLD film growing in the range of 600 to 925°C could change its surface morphology over time and annealing temperatures above 800°C. For such reasons the anatase film synthesis involves two steps: growth at a given temperature plus a long annealing in air or oxygen at high temperatures, e.g. 900°C. These steps allow the anatase islands to coalesce into flat terraces (Figure 7). This result also offers a host temperature range to grow anatase films by MBE at temperatures equal or higher than 800°C since the island coalescence process is already activated in the vicinity of this temperature [28].

The two-step growth method (film growth and post-annealing treatment in oxygen or air at ambient pressure) used during the synthesis of anatase films by the PLD and MBE techniques, produces films of similar crystalline quality (section 3.4) and show between them a very similar lattice strain response to STO(001) mismatch. The lattice constants obtained from synchrotron diffraction experiments [29] for these films after averaging several reflections to compensate miscut misalignments along the surface normal direction are:

PLD: (H,K,L = 0,1,1)$_{film}$ = (0,1.0335, 0.4091)$_{STO}$ : (a$_{film}$, c$_{norm}$)$_{PLD}$ = (3.7784$_5$, 1.0033$_8$)
MBE: (H,K,L = 0,2,4)$_{film}$ = (0, 2*1.0328, 1.6397)$_{STO}$ : (a$_{film}$, c$_{norm}$)$_{MBE}$ = (3.7811$_5$, 1.0013$_8$)
which are very close but not identical probably due to their different film thicknesses.

The preparation methods by PLD and MBE shown in Figures 8 and 10 generate patterns of coalescence of islands like those observed by other authors [28,34] with similar sizes, however, there are some differences between them. The 10 nm film synthesized by PLD at 800°C [34] indicates that island coalescence is reached at the temperature of growth, however, the work does not offer any information on whether or not any post-heat treatment was applied to the film. In our 50 nm anatase thin films (Table 2), coalescence is always achieved after 6 hours of annealing at 750-800 °C or higher (Figure 7). The samples prepared at any growth temperature by PLD, without post annealing, never coalesce as evidenced in Figure 6. Consequently, annealing of the film at high temperatures (and in oxygen to avoid its deficiency) is mandatory to induce coalescence of the grains or islands. The large domain sizes achieved by MBE at a deposition temperature of 750°C [28] are recovered in the present films after a long air annealing at high temperatures, although it is true that we have not carried out long anneals at 750°C because in our films the higher the annealing temperature is the larger the terraces are. The largest terrace size achieved in this work has been reached at an annealing temperature of 925 °C (Figure 8). In accordance with this trend, the effect of annealing closer to the anatase/rutile transformation temperature, detected at 950 °C, would further increase the size of the terraces.

Coalescence results in the melting of the molten islands to form larger terraces on the film surface. This process enhances atomic diffusion and consequently reduces the concentration of holes or gaps observed between neighbouring islands (Figure 6). However, these holes do not disappear completely. This atomic rearrangement induced during the melting process of the islands makes them larger, while also producing a rearrangement of the distribution of the holes, reducing their number but making them larger. These larger holes would presumably be formed by flat terraces of anatase towards deeper levels [28], although the holes could also reach the substrate surface (Supplementary information, section I). The maximum hole depth detected is close to 15 nm, as shown in Figure 8, after long anneals at high T. The larger number of "hole-defects" in the GT films would result in a higher concentration of $Ti^{3+}$ than the annealed samples, which would presumably make them more interesting for catalysis [29].

The anatase to rutile transformation begins to appear at 950ºC (Figure S4) but the anatase phase is still the majority phase present in the film at 1000ºC (Figure 5). This window of temperatures allows the possibility of controlling the percentages of each phase in the film for specific purposes, for example, to study effects of the phase ratio on the properties of these catalysts…, since the anatase to rutile transformation is progressive with T in these films. Even at higher temperatures, such as 1100°C, complete conversion of anatase to rutile is still not achieved [65 & ref. therein].

The quality of the anatase (001) MBE and PLD films was tested by measuring the (4x1) surface reconstruction using STM techniques, as mentioned in section 3.4. The average sizes of the crystalline domains for both types of films were on average 120 nm. This size is still insufficient to use these samples for surface structure determination. However, the average terrace sizes of 600 nm obtained after annealing at high T for extended times (Figure 8) are promising candidates for surface X-ray diffraction experiments. So, the proposed annealing process could help to overcome this problem.

Another aspect to be investigated is the quality of MBE growth of these films by introducing monoatomic oxygen into the UHV chamber instead of gaseous molecular oxygen. In this case, the higher reactivity of monoatomic oxygen would accelerate the growth process and reduce the evaporation time.

In summary, as mentioned throughout the article, PLD and MBE techniques can share common steps that facilitate the film growth process: 1) surface preparation of the substrates by ultrasonic baths in deionized water and annealing at thigh temperature, 2) introduction into their corresponding chambers and cleaning of the carbon substrate surface by annealing at moderate temperature under oxygen pressure and 3) post annealing heat treatments after the film growth process at high temperatures under ambient or oxygen pressure.

**CONCLUSIONS**

Systematic growth of epitaxial anatase (001) films by PLD and MBE techniques shows similar film qualities after annealing for several hours in oxygen-rich atmospheres. The effects of temperature and time on film growth and the role of these two parameters during the film annealing process have been studied independently. The post-growth annealing process enlarges the size of the ordered surface terraces in 50 nm thick films to 400 nm as a consequence of the coalescence of the grains into flat terraces after 6 hours of annealing at 900 ºC.

In addition, it is possible to control the average lateral size of the terraces or domains of the anatase (001) surface, almost independently of the growth temperature, by performing a long post-growth anneal at high temperatures (900-925°C for 6-12 hours) in air or oxygen. A larger increase in terrace size would be expected as a result of island coalescence from larger grain sizes, such as the films synthesized at 925°C (Figure 6G). In fact, a 12-hour long anneal at 925°C on such a sample increases the surface terrace dimensions up to 650 nm (Figure 8), highlighting the relevance of the combined effects of temperature and time on each other in these films. Experiments performed near the anatase-to-rutile temperature transformation (950°C-1000°C) offer a possible temperature window between 925°C and 950°C to optimize the recrystallisation of anatase films before the anatase-to-rutile transformation occurs.

The evolution of the in-plane and out-of-plane lattice parameters with temperature has been described for freshly grown anatase films. The out-of-plane lattice parameter evolves from a compressive to a tensile regime with increasing temperature, while the in-plane lattice parameters remain under a compressive regime in the same temperature range explored. The biaxial compressive strain behaviour of the in-plane film lattice parameters induces a reduction of the cell volume over the whole temperature range explored.

The evolution of the positions, widths and intensities of the anatase XRD peaks of the grown and annealed films with temperature indicates that two different deformation strain contributions have to be considered: one generates the strain that causes the continuous expansion of the c-lattice parameter ($\varepsilon_{Lat}$), the other ($\varepsilon_{mic}$) is temperature dependent and reduces the defects and microdistortions present in the films during the annealing process. The reduction of microstructural strain comes at the cost of increased lattice strain in case the anneals are performed at temperatures above GT. The continuous increase of the c-lattice parameter with T, annealing or GT, suggests that the deformation associated with it is independent of the level of defects in the films and its dependence is mainly established through T, while the microstructure seems to be related mainly to the lattice mismatch as it does not evolve with GT.

In both types of films, PLD and MBE, the 4x1/1x4 anatase surface reconstruction is observed in UHV conditions after two successive cycles of Ar$^+$ ion sputtering (0.6 keV, $P_{Ar} = 5*10^{-6}$ mbar, t=5min) and annealing at 850°C under an oxygen pressure of $10^{-6}$ mbar.

## ACKNOWLEDGEMENTS

The following funding is acknowledged: grant PID2021-123276OB-I00 funded by MCIN/AEI/10.13039/501100011033 and by "ERDF A way of making Europe"; Severo Ochoa Programme for Centre of Excellence in R&D, FUNFUTURE (CEX2019-000917-S) funded by MCIN/AEI /10.13039/501100011033; grant LINKA20394 funded by CSIC, program i-LINK 2021; synchrotron experiment was awarded by SOLEIL and performed on the CRG IF beamline at ESRF, France (proposal number 20200213). J.M.C. also thanks Severo Ochoa Programme (Grant CEX2021-001214-S funded by MCIN/AEI/10.13039/501100011033).## REFERENCES